# Photocarrier Injection and Current-Voltage Characteristics of a La$_{0.8}$Sr$_{0.2}$MnO$_3$/SrTiO$_3$:Nb Heterojunction at Low Temperature

Takaki Muramatsu, Yuji Muraoka and Zenji Hiroi[*]

*Institute for Solid State Physics, University of Tokyo, Kashiwa, Chiba 277-8581, Japan*

An epitaxial heterojunction made of a *p*-type perovskite manganite La$_{0.8}$Sr$_{0.2}$MnO$_3$ film and an *n*-type strontium titanate SrTiO$_3$:Nb substrate has been fabricated by the pulsed laser deposition technique. The $I - V$ characteristics and photovoltaic properties are measured under a UV light irradiation in a wide temperature range down to 10 K. It is found that the junction works as an efficient UV photodiode with a large external quantum efficiency of 28 %. It is also demonstrated that the manganite film can be doped with a certain amount of holes by the photocarrier injection. The maximum surface hole density is estimated to be $3.0 \times 10^{13}$ cm$^{-2}$.

---

[*] To whom correspondence should be addressed to E-mail: muramatu@issp.u-tokyo.ac.jp

## 1. Introduction

Solid state devices made of transition metal oxides (TMOs) are of great interest, because various TMOs exhibit interesting properties such as metallic ferromagnetism or high-$T_c$ superconductivity. These properties in TMOs are realized by doping carriers to parent Mott insulators in which electrons are localized due to a large Coulomb repulsion. Usually, carrier doping is achieved by the chemical substitution of constituent elements, which may give rise to local lattice distortions or randomness. Therefore, there have been efforts to achieve an alternative carrier control using conventional semiconductor techniques in oxide heterostructures. For example, carrier doping using a field-effect transistor (FET) has been extensively studied. In the case of a $La_{0.8}Sr_{0.2}MnO_3$ (LSMO) film grown on an insulating $SrTiO_3$ (STO) substrate, Katsu *et al.* observed a decrease in the Curie temperature by 10 K under ultraviolet (UV) light irradiation.[1] They suggested that electrons generated in the substrate were injected into the film.

Recently, Muraoka *et al.* reported a novel method for hole doping called the photocarrier injection (PCI) method in an oxide heterostructure of $VO_2$/$TiO_2$:Nb under UV light irradiation.[2] The PCI method was also applied to other TMOs like manganese oxides, cuprates and organics, demonstrating the generality of this method.[3-10] In a high-$T_c$ cupric oxide heterostructure of $YBa_2Cu_3O_{7-\delta}$/STO:Nb, the increase of $T_c$ by 5 K was found as a result of hole injection under UV light irradiation.[6] In the case of a LSMO/STO:Nb heterojunction, a relatively small photovoltage of 0.09 V was observed at room temperature, compared with 0.4 V found in $VO_2$/$TiO_2$:Nb and 0.8 V in $YBa_2Cu_3O_{7-\delta}$/STO:Nb, which may reflect the differences in the band structure of the heterojunctions.

A simple band picture to explain the PCI has been proposed.[5] Nb doped STO is an *n*-type degenerate semiconductor with a wide bandgap of 3.2 eV, while most TMOs are *p*-type with smaller bandgap of about 1.0 eV. Thus, the band structure of the heterojunction must be essentially similar to that of an anisotype heterojunction with a depletion layer at the interface. The $Nb^{5+}$ donor level of STO:Nb is expected to be located just below the bottom of the conduction band. The Fermi level of STO:Nb is located above that of most TMOs, with the difference between them giving rise to a potential barrier of a height $V_d$ and a depletion layer of a width $w$ in the substrate. When a photon with energy larger than the bandgap of STO is absorbed to create an electron-hole pair in the substrate, it is expected from the band diagram that only the hole can be injected into the film with the electron remaining in the substrate. Since the hole has much larger potential energy in STO:Nb than in TMO, this drift process must proceed quickly. In contrast, the recombination between the hole in the film and the electron in the substrate must be slow, because it occurs by tunneling across the barrier. As a result, the lifetime of the injected hole can be sufficiently long to increase substantially the hole carrier density in the TMO film.

In this report, we study the LSMO/STO:Nb heterostructure in more detail in terms of both PCI and a UV sensitive photodiode. The PCI efficiency is estimated in a wide temperature range down to 10 K by calculating injected hole density in the LSMO film.

## 2. Experimental

A heterostructure was fabricated by epitaxial growing a LSMO thin film on a 0.05 wt% Nb-doped STO (100) single crystal substrate (Earth chemical) by the pulsed laser deposition (PLD) technique. First the STO:Nb substrate was annealed under high vacuum ($P < 10^{-6}$ Pa) at 750-800℃ for 30 minutes in order to evaporate impurities on the surface, and then a LSMO film was deposited on the substrate at temperature of 750 ~ 800   in an oxygen partial pressure of 10 Pa. A typical film thickness was 50 nm as analyzed by observing the Laue oscillation of diffracted x-ray intensity in a diffractometer (X'Pert MRD, Philips). A sintered LSMO polycrystalline pellet was used as a target. A 100 W xenon lamp (Asahi Spectra Co.,Ltd.) with a variable light intensity filter was used as a light source with $\lambda$ = 300 ~ 400 nm.. A wide range of light intensity of $10^{-5} \sim 10^2$ mWcm$^{-2}$ was attained by the combination of two neutral density filters. Junction capacitance was obtained by measuring AC impedance, using a HP-4284A precision LCR-meter working at a frequency of 20 Hz and an amplitude of 20 mV. Current density-voltage (*I-V*) characteristics and photovoltaic effects were measured by using a Keithley 2400 source meter and an Agilent Technology 34401A digital multimeter. Low-temperature experiments down to 10 K were carried out in an optical cryostat equipped with a Gifford-McMahon refrigerator and a quartz window for UV light illumination.

Electrodes were selected carefully in order to attain ohmic contact in a wide temperature range from room temperature to 10 K. They were a thermosetting silver paste on the LSMO film and a metal composite on the STO:Nb substrate. The latter was made by sputtering a titanium metal film under high vacuum of $10^{-6}$ Pa and then depositing a gold film on it to avoid oxidation. The alignment of the electrodes on the sample is shown in the inset to Fig. 1.

**3. Results and discussion**

3.1 *I - V* characteristics

Figure 1 shows the *I - V* characteristics of the LSMO/STO:Nb junction measured in the dark and under UV light irradiation with a light intensity of 80 mW/cm$^2$ at various temperatures. Rectifying behavior is clearly observed in the dark. The *I - V* curves shift downward almost in parallel, when irradiated by a UV light as in a conventional photodiode. The effective diffusion voltage $V_d$ is estimated by fitting the *I-V* curves assuming an equivalent circuit as reported previously.[6] As shown in Fig. 2(a), $V_d$ increases linearly with decreasing temperature. Thus, the intrinsic diffusion voltage is determined to be 0.65 V by extrapolating the curve to 0 K.

A feature to be noted in Fig. 1 is the large bias voltage dependence in the third quadrant. The leakage current increases with increasing negative bias voltage. Especially at 300 K and 200 K the leakage current increases largely below -0.5 V, indicating the beginning of the reverse breakdown as often observed in a conventional semiconductor *pn* junction. In contrast at 100 K and 10 K it becomes insensitive to bias voltage. This reverse breakdown may originate from the Zener effect which is caused by tunnel current through a potential barrier at the interface.

In order to attain information on the band diagram, we have analyzed leakage current $I_\text{L}$ as functions of temperature and bias voltage. The temperature dependence of $I_\text{L}$ measured at several bias voltages in the dark is plotted in Fig. 2(b). At a large bias voltage between -1.0 V and -0.8 V, the $I_\text{L}$ shows almost linear behavior in the semi-logarithmic plot. On the other hand, at bias voltages between -0.4 V and -0.1 V, the $I_\text{L}$ decreases more rapidly with decreasing temperature, and becomes almost constant below 100 K. The former is ascribed to the Zener effect, while the latter is due to thermally excited carriers at high temperature. The crossover voltage between the two regimes should correspond to the energy difference between the top of the valence band of the LSMO film and the bottom of the conduction band of the STO:Nb substrate. In fact it is close to the value of $E_\text{g}/e - V_\text{d} = 1.1 \text{ V} - 0.05 \text{ V} = 0.45 \text{ V}$.

A probable energy diagram for the present junction is depicted in Fig. 3. The Fermi energy $E_\text{F}$ of STO:Nb must be located just above the bottom of the conduction band, because it is a degenerate semiconductor. On the other hand, the above estimations suggest that the $E_\text{F}$ of LSMO lies slightly below the middle of the bandgap in spite that it is a *p*-type metal. Presumably, a conventional band picture may not be applicable to LSMO, because the origin of the bandgap is essentially electron correlation effects. Our results suggest that the Fermi level of LSMO is located at a mid-gap state which has been produced as a result of hole doping into a Mott insulator, as in the case of copper oxide superconductors.[11, 12]

The temperature dependence of short-circuit current $I_{sc}$ is shown in Fig. 2(c), which exhibits a broad maximum at 100 – 200 K. The $I_{sc}$ is given by the equation $I_{sc}/e = (L/h\nu)\gamma_{ext}$, where $L$, $h\nu$ and $\gamma_{ext}$ are UV light intensity, photon energy and external quantum efficiency, respectively. Then, $\gamma_{ext}$ = 28 % at 100 K and 11 % at 300 K, assuming that the incident light has a wavelength of 350 nm. These values are sufficiently large compared with other sophisticated UV photodetectors made of wide-gap semiconductors such as (Al,Ga)N.[13]

3.2 Photovoltaic measurements

We have carried out photovoltaic measurements in order to estimate injected hole carrier density. The light irradiation dependence of $V_{oc}$ at various temperatures from 300 K to 10 K is shown in Fig. 4. At 300 K a positive $V_{oc}$ to the LSMO film increases linearly with increasing light intensity in a wide range of $10^{-4} < L < 10$ mW/cm$^2$, tends to deviate from the linear behavior above 10 mW/cm$^2$, and saturates at 80 mW/cm$^2$ with the maximum value of 0.08 V. This is because hole doping must proceed in a rigid band at low irradiance of $V_{oc} \ll V_d$, while at high irradiance an effective Schottky barrier would be reduced to $V_d - V_{oc}$, resulting in the observed saturation of $V_{oc}$. At 10 K in the low intensity region the $V_{oc}$ is largely enhanced by a factor of 100 compared with that at room temperature. In the high intensity region the enhancement becomes smaller, about five times larger than at room temperature. This enhancement is due to the improvement of the efficiency of PCI at low temperature.

3.3 $C$ - $V$ measurements

A photodiode is considered as a condenser to accumulate photogenerated charges. The junction capacitance $C$ has been measured as a function of bias voltage, and is shown in the $C^{-2}$ - $V$ plot of Fig. 5. In a Schottky-type junction, the junction capacitance is expressed by an equation $C^{-2} = 2(V_d-V)/A^2eN_D\varepsilon_0\varepsilon_s$, where $N_D$, $V_d$, $A$, $\varepsilon_0$ and $\varepsilon_s$ are donner concentration, diffusion voltage, effective area of junction, vacuum permittivity and relative permittivity constant, respectively. Then, $C^{-2}$ should be proportional to the bias voltage $V$. In Fig.5 the $C^{-2}$ - $V$ curve at 300 K shows an only small linear region for $V$ = -0.35 V ~ -0.15 V. This is due to a leak current by thermally excited carriers as observed in the $I$ - $V$ characteristics of Fig.1. The linear region is widened as temperature decreases, because the leak current is suppressed.

The intrinsic diffusion voltage can be estimated from the 10 K data to be approximately 0.5 V by the extrapolation of the $C^{-2}$ - $V$ curve to the $V$ axis. This value is close to $V_d$ = 0.65 V attained from the $I$ - $V$ characteristics. The $\varepsilon_s$ value of the depletion layer is also evaluated from the slope of the linear $C^{-2}$ - $V$ plot. The temperature dependence of $\varepsilon_s$ thus determined is shown in Fig. 6. The $\varepsilon_s$ is 350 at 250 K, which is close to the bulk value of 330[14], and is almost temperature independent above 100 K. However, the $\varepsilon_s$ decreases to 120 at 10 K. This is in contrast to the case of pure STO which exhibits a large enhancement in $\varepsilon_s$ at low temperatures below 100 K due to quantum paraelectric. Probably, niobium doping has suppressed the enhancement in our STO:Nb substrate.

3.4 Estimation of carrier density

The surface density of injected holes $Q/e$ is calculated from a simple relation of $dQ = CdV$ by integrating the $C$ over $V_{oc}$ using the $\varepsilon_s$ value from the $C^{-2}$ - $V$ measurements. The obtained $Q/e$ is plotted in Fig. 7 as functions of $L$ and $T$. At 300 K the doped maximum surface hole density is $8.3 \times 10^{12}$ cm$^{-2}$. Assuming a uniform distribution in the film of 10 nm thick, the hole concentration would be 0.06 % per Mn. At 100 K the maximum hole surface density rises up to $3.0 \times 10^{13}$ cm$^{-2}$, corresponding to 0.2 % per Mn. The $Q/e$ decreases at lower temperature, which is ascribed to the decrease of $\varepsilon_s$. The maximum value attained at 100 K is apparently too small to control the physical properties of the LSMO film. However, injected holes are expected to be located near the interface by an attractive force from electrons left in the substrate, and are possibly accumulated with higher density near the interface. If one assume a one-nm thick active layer, the maximum hole density would be 2 % per Mn, which is still not so large to expect a large change in the properties. Actually we could not detect a significant change in the curie temperature of the LSMO film by magnetic measurements ascribed to the PCI.

A key to further enhance the efficiency of PCI and inject a sufficiently large number of holes to control the properties of LSMO is to increase the $\varepsilon_s$ of substrate. The use of an alternative substrate such as $(Sr,Ba)TiO_3$ would be helpful.

**4. Conclusion**

A high quality heterojunction of LSMO/STO:Nb has been fabricated, and its response to a UV light is investigated down to low temperature. It is found that the junction works as an efficient UV photodiode with an external quantum efficiency of 28 %. The maximum surface hole density injected to the LSMO film is estimated to be $3.0 \times 10^{13}$ cm$^{-2}$.

Acknowledgements

We thank T. Yamauchi for his help in electrical measurements and J. Yamaura for helpful discussion. This work is supported by a Grant-in-Aid for Creative Scientific Research (IJNP0201) and a Grant-in-Aid for Scientific Research (14750549) given by the Ministry of Education, Culture, Sports, Science and Technology.

Fig. 1.

*I-V* characteristics of a $La_{0.8}Sr_{0.2}MnO_3/SrTiO_3$:Nb junction measured at various temperatures in the dark and under UV light irradiation with a light power of 80mW/cm$^2$. Inset shows the schematic configuration of the sample and electrodes.

Fig. 2.

(a) Temperature dependences of diffusion voltage $V_d$ obtained from fitting the *I - V* curves shown in Fig. 1 (circle) and the maximum open-circuit voltage $V_{oc}$ (triangle). (b) Temperature dependence of leakage current density $I_L$ at various negative bias voltages. (c) Temperature dependence of short circuit current density at $L = 80$ mW/cm$^2$.

Fig .3.

Schematic band diagram of the $La_{0.8}Sr_{0.2}MnO_3/SrTiO_3$:Nb heterojunction.

Fig. 4.

UV light irradiance dependence of open-circuit photovoltage on $La_{0.8}Sr_{0.2}MnO_3/SrTiO_3$:Nb.

Fig. 5.

Bias voltage dependence of $C^{-2}$ on $La_{0.8}Sr_{0.2}MnO_3/SrTiO_3$:Nb at several temperatures measured in the dark with an ac frequency of 20 Hz and the amplitude of 20mV.

Fig. 6.

Temperature dependence of relative permittivity estimated from the $C^{-2} - V$ of relation of Fig. 5.

Fig. 7.

(a) UV light irradiance dependence of surface hole density on $La_{0.8}Sr_{0.2}MnO_3/SrTiO_3{:}Nb$.

(b) Temperature dependence of the maximum surface hole density on $La_{0.8}Sr_{0.2}MnO_3/SrTiO_3{:}Nb$.

Fig. 1

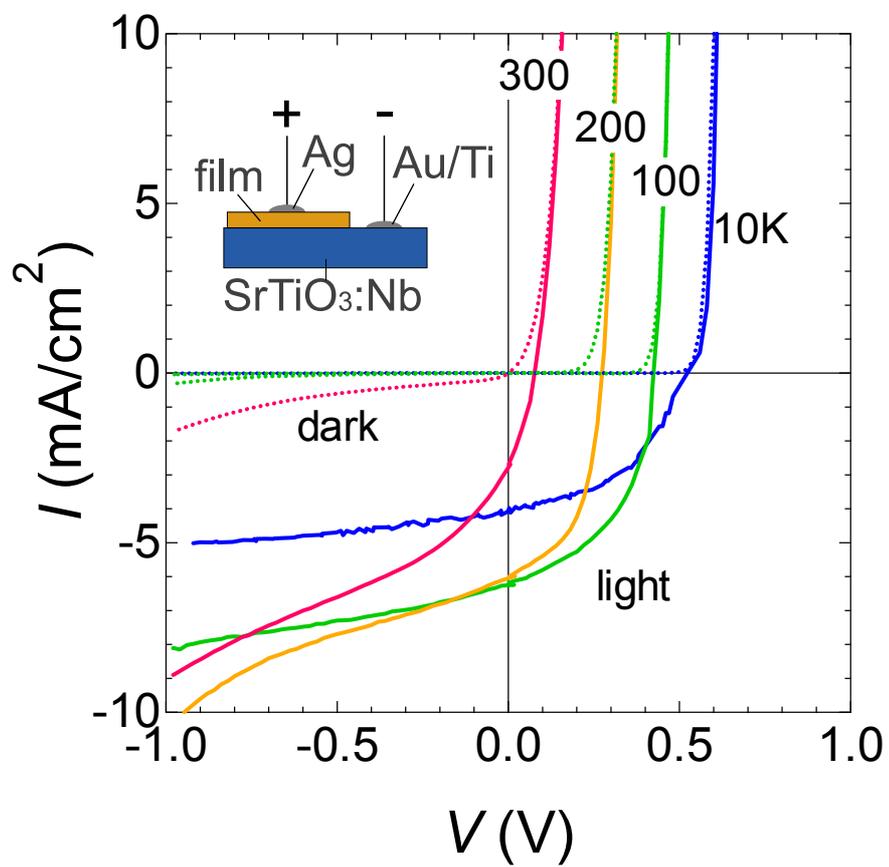

Fig. 2

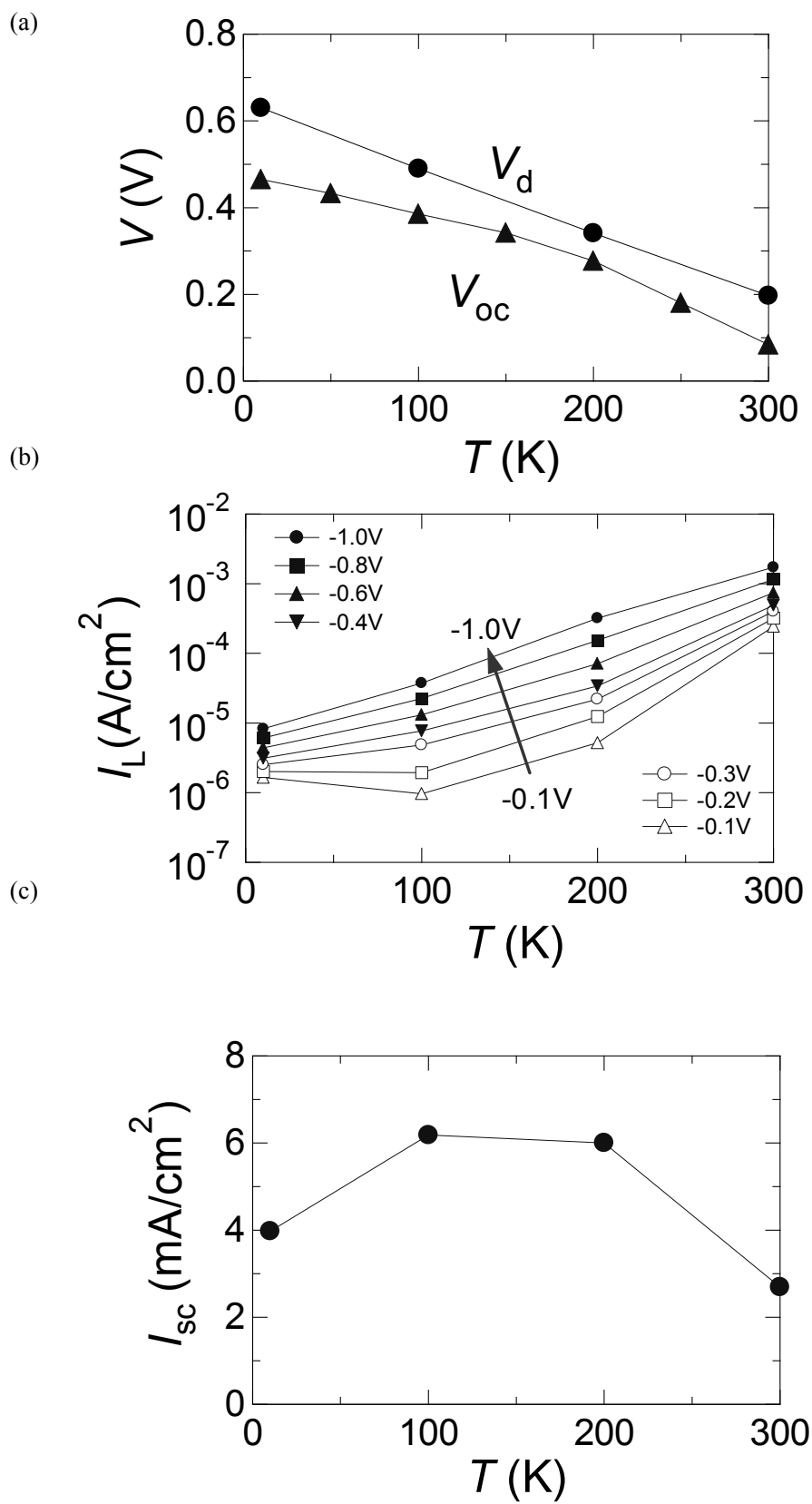

Fig. 3

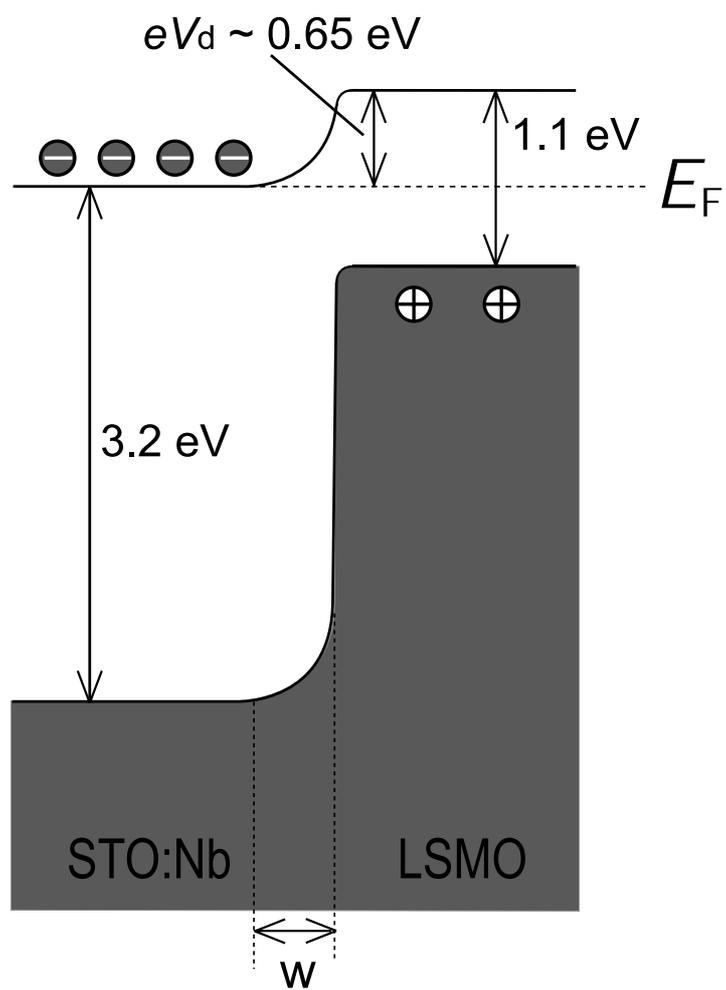

Fig. 4

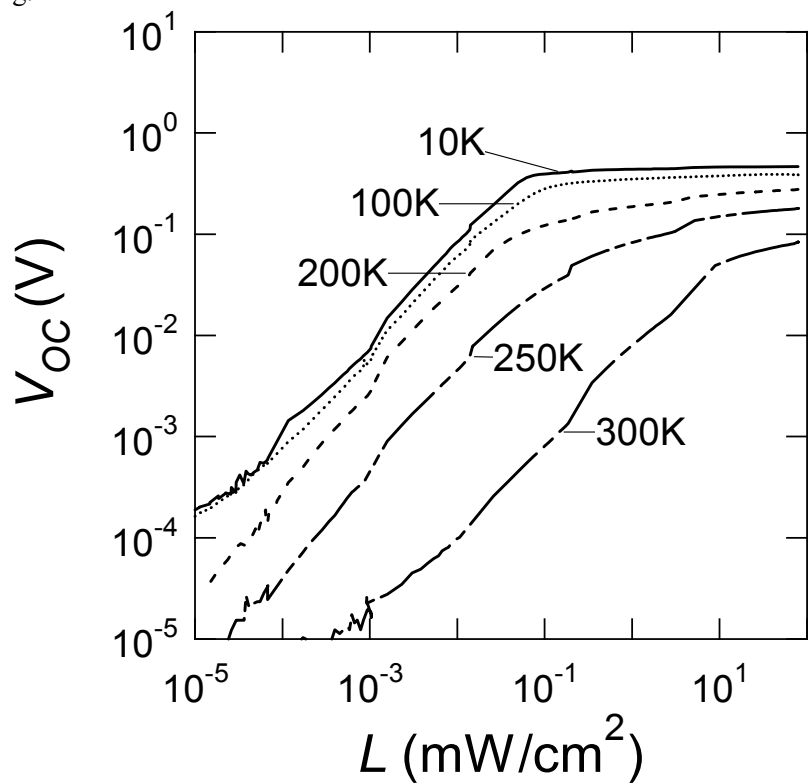

Fig. 5

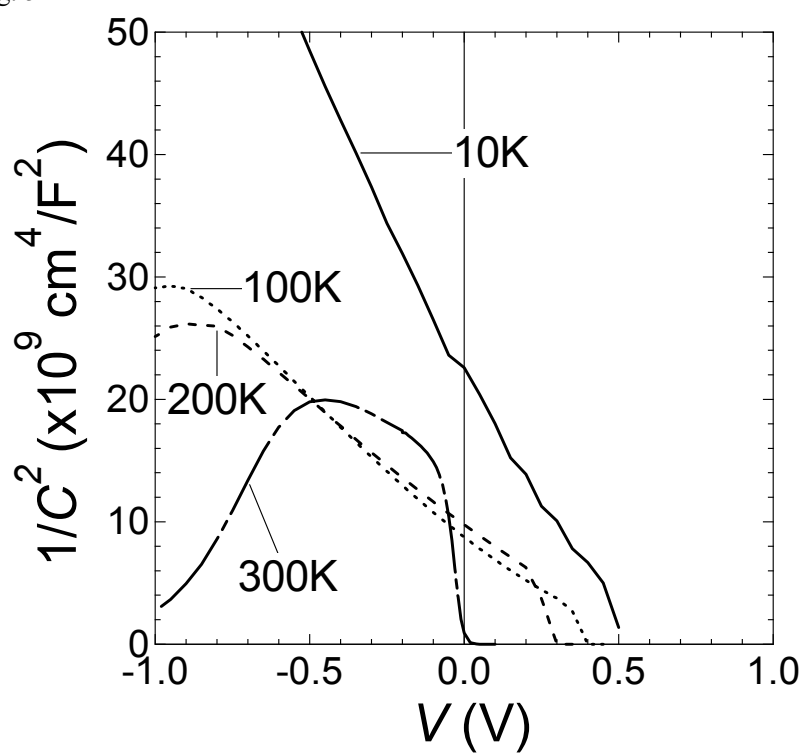

Fig.6

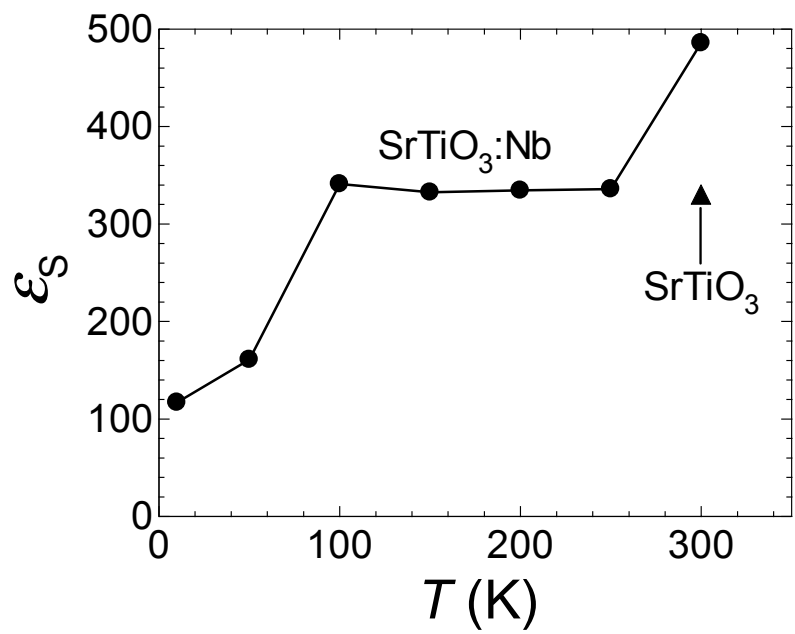

Fig. 7
(a)

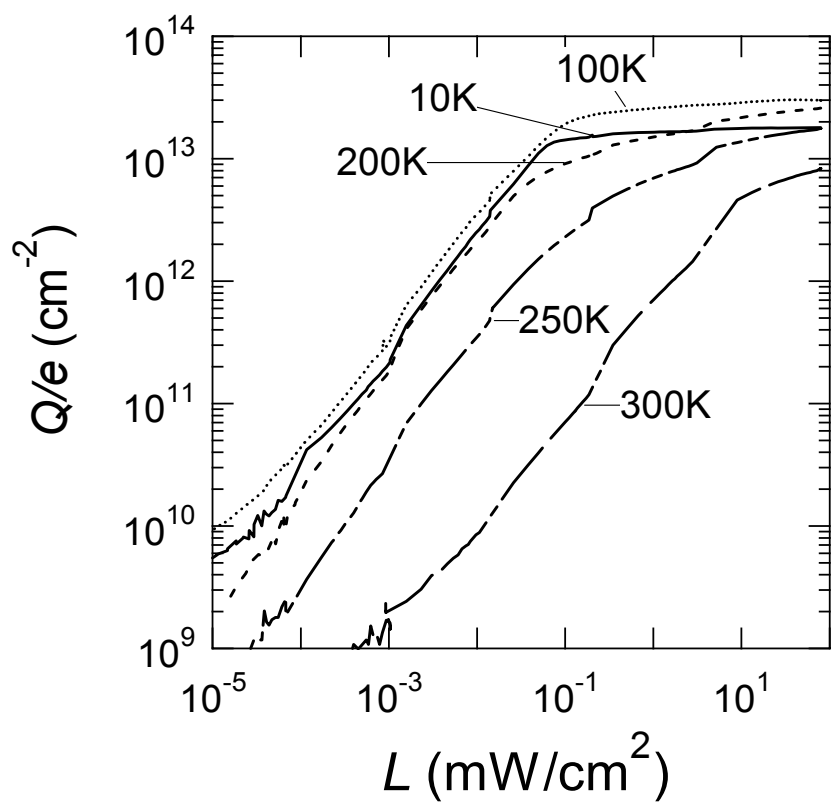

(b)

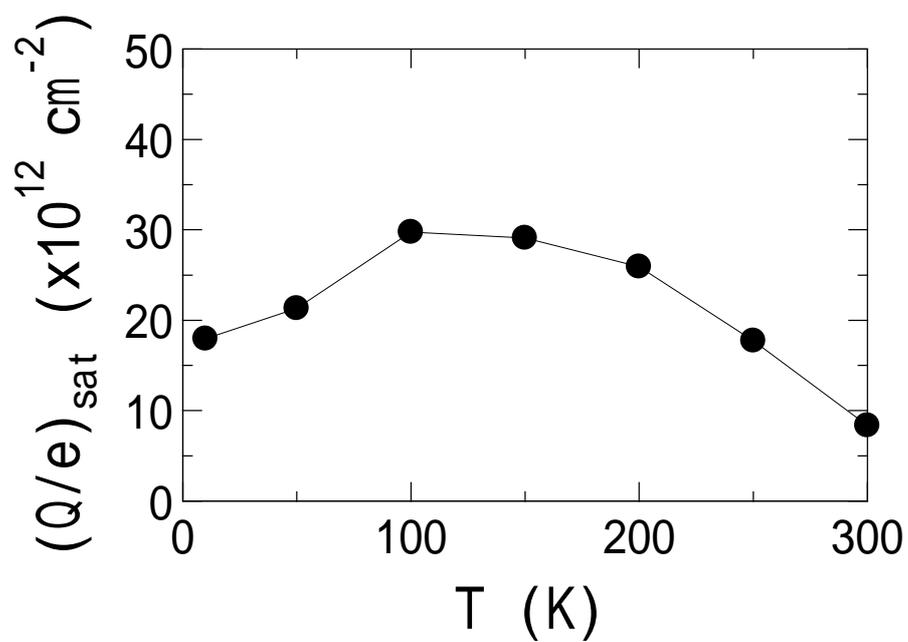